\documentclass[aps,prd,reprint,longbibliography,nofootinbib,superscriptaddress,floatfix]{revtex4-2}

\usepackage[english]{babel}
\usepackage[utf8]{inputenc}
\usepackage{amsmath}
\usepackage{mathbbol}
\usepackage{amssymb}
\usepackage{bbold}
\usepackage{graphicx,amsfonts}
\usepackage{epsfig}
\usepackage{hyperref}
\usepackage{bm}
\usepackage{mathrsfs}
\usepackage{mathtools}
\usepackage{enumerate}
\usepackage{amsthm}
\usepackage{bbm}
\usepackage{comment}
\usepackage{physics}
\usepackage{url}
\usepackage{upgreek}
\usepackage[svgnames]{xcolor}
\usepackage[title]{appendix}
\definecolor{teal}{RGB}{0, 128, 128}
\definecolor{myred}{RGB}{179, 27, 27}
\hypersetup{
	colorlinks=true,
	linkcolor=teal, 
	citecolor=teal, 
	urlcolor=teal  
}

\begin{document}
\title{Spectral instability of horizonless compact objects within astrophysical environments}
\author{Kyriakos Destounis}
\email{kyriakosdestounis@tecnico.ulisboa.pt}
\affiliation{CENTRA, Departamento de Física, Instituto Superior Técnico – IST, Universidade
de Lisboa – UL, Avenida Rovisco Pais 1, 1049-001 Lisboa, Portugal}
\author{Mateus Malato Corrêa}
\affiliation{Programa de Pós-Graduação em Física, da Universidade Federal do Pará, 66075-110, Belém, PA, Brazil}
\author{Caio F. B. Macedo}
\affiliation{Faculdade de Física, Campus Salinópolis, Universidade Federal do Pará, 68721-000, Salinópolis, Pará, Brazil}
\author{Rodrigo Panosso Macedo}
\affiliation{Center of Gravity, Niels Bohr Institute, Blegdamsvej 17, 2100 Copenhagen, Denmark}
%
\begin{abstract}
Recent non-modal analyses have uncovered spectral instabilities in the quasinormal-mode spectrum of black holes; a phenomenon that intriguingly extends to spherically-symmetric exotic compact objects. These results point to a sensitivity of the spectrum with potentially far-reaching implications for black-hole spectroscopy. At the same time, growing attention has turned to astrophysical environments around compact objects and their role in shaping gravitational-wave astrophysics. In this work, we establish a direct link between spectral instabilities and environmental effects by modeling matter as a localized bump outside the light ring of a spectrally-unstable exotic compact object with a purely reflective surface. We find that while such environments can destabilize the fundamental quasinormal modes of loosely-compact exotic objects, the fundamental modes of ultra-compact horizonless objects remain remarkably robust. In contrast, overtones are shown to develop spectral instabilities in the presence of the bump. By tracking both interior modes, trapped between the light ring and the surface of the exotic compact object, and exterior modes, confined between the bump and the light ring, we uncover an overtaking instability in which ``unperturbed'' exterior overtones metamorphose into ``perturbed'' fundamental modes as the bump moves outward. Finally, we demonstrate that environmental effects, while capable of further amplifying spectral instabilities, cannot induce next-to-leading-order perturbations strong enough to trigger a modal instability.
\end{abstract}

\maketitle

\section{Introduction}

Gravitational-wave (GW) astrophysics has entered a transformative era since the landmark detection by LIGO in 2015 \cite{LIGOScientific:2016aoc}, followed by the most recent and loudest event in 2025 \cite{Others:2025nbi}, thus confirming a century-old prediction of Einstein’s General Relativity (GR). Current state-of-the-art efforts involve a global network of detectors, including LIGO, Virgo, and KAGRA, operating with ever-increasing sensitivity to capture faint ripples in spacetime caused by cataclysmic astrophysical events like black hole (BH) and neutron star mergers \cite{LIGOScientific:2025slb}. These observations have opened a new window into the universe, enabling precision tests of GR in extreme regimes and offering insights into the population, formation, and evolution of compact objects \cite{Barack:2018yly,Buoninfante:2024oxl}. Meanwhile, next-generation, observatories, such as the Laser Interferometer Space Antenna \cite{LISA:2017pwj}, the Einstein Telescope \cite{Punturo:2010zz,Hild:2010id,ET:2019dnz}, and Cosmic Explorer \cite{Reitze:2019iox,Evans:2021gyd} among others, promise to expand the observable sensitivity and frequency range onto uncharted regions of the Cosmos \cite{LIGOScientific:2017ync,Barausse:2020rsu,Amaro-Seoane:2022rxf,LISA:2022yao,LISA:2022kgy,Karnesis:2022vdp,Ezquiaga:2020tns,Hall:2022dik,Abac:2025saz}. Thus, they will provide access deeper into the early universe \cite{LISACosmologyWorkingGroup:2022jok,You:2020wju,Cai:2016sby,Belgacem:2019tbw}, uncover beyond the Standard Model physics, such as dark matter \cite{Cerdonio:2008un,Macedo:2013qea,Cannizzaro:2022xyw,Bringmann:2023iuz}, perceive the role of astrophysical environments around BHs \cite{Barausse:2014tra,Destounis:2020kss,Cardoso:2020nst,Destounis:2021mqv,Destounis:2021rko,Cardoso:2021wlq,Vicente:2022ivh,Cardoso:2022whc,Cardoso:2020lxx,Destounis:2022obl,LimaJunior:2022gko,Polcar:2022bwv,Cannizzaro:2024yee,Eleni:2024fgs,Mollicone:2024lxy,Pezzella:2024tkf,Macedo:2024qky,Speeney:2024mas,Duque:2023seg,Mitra:2025tag,Gliorio:2025cbh,Ovgun:2025bol,Chowdhury:2025tpt,Fernandes:2025lon,Vicente:2025gsg,Zwick:2025ine,Spieksma:2024voy,Li:2025zgo,Kouniatalis:2025itj} and unearth dark exotic compact objects (ECOs) that are currently theorized \cite{Kesden:2004qx,Cardoso:2007az,Cardoso:2016rao,Cardoso:2016oxy,Cardoso:2017cqb,Maggio:2020jml,Cardoso:2019rvt,Cardoso:2021ehg,Destounis:2023khj,Destounis:2023gpw,Rosato:2025byu}. The pay-off for facing these outstanding issues, regarding the existence of ECOs, is to be able to quantify the statement that BHs exist in nature.

GWs carry fundamental information regarding the early, intermediate and late stage of binary coalescence. The merger remnant forms after the inspiral of the two objects, that collide violently while releasing the remaining ``whispers'' of the binary, i.e. the ringdown, that eventually relaxes to a remnant in equilibrium. Analyzing the ringdown and its respective, damped-sinusoid oscillation modes provides the backbone of the \emph{BH spectroscopy program} \cite{Dreyer:2003bv,Giesler:2019uxc,Destounis:2023ruj,Carullo:2025oms,Berti:2025hly}, which is based on decomposing the ringdown signal with respect to an infinite set of discrete quasinormal modes (QNMs) \cite{Kokkotas:1999bd,Berti:2009kk,Konoplya:2011qq}. 

The landscape of QNMs is currently expansive. A key objective in this domain is to leverage the QNMs of BHs and other compact objects in order to test GR at the phenomenological \cite{Leung:1997was,Berti:2004md,Berti:2015itd,Glampedakis:2017dvb,Kobayashi:2012kh,Kobayashi:2014wsa,Berti:2019xgr,Cardoso:2019mqo,McManus:2019ulj,Kimura:2020mrh,Hirano:2024fgp,Franchini:2023eda,Chen:2021cts,Baibhav:2023clw,Berti:2025,LIGOScientific:2025obp} and fundamental level \cite{Dafermos:2003wr,Dafermos:2012np,Dafermos:2014cua,Dafermos:2021cbw,Hintz:2015jkj,Hintz:2015koq,Hintz:2016gwb,Hintz:2016jak,Hintz:2025hvr,Cardoso:2017soq,Cardoso:2018nvb,Destounis:2018qnb,Liu:2019lon,Destounis:2019omd,Destounis:2020yav,Dias:2018ufh,Mo:2018nnu,Dias:2018ynt,Casals:2020uxa,Davey:2024xvd,Steinhauer:2025bbs}. Notably, the stark contrast between the QNM spectra of BHs and ECOs, driven by drastic changes in boundary conditions, could serve as a signature for new physics at the horizon scale \cite{Foit:2016uxn,Konoplya:2018yrp,Cardoso:2019apo,Coates:2019bun,Agullo:2020hxe,Coates:2021dlg,Chakraborty:2022zlq}. However, GW data are inherently recorded in the time domain. Present detectors can only resolve the early-time ``prompt ringdown,'' primarily associated with perturbations of the light ring. As a result, BH spectroscopy is currently limited to this brief, high-frequency phase of the observed ringdown \cite{Cardoso:2016rao,Cardoso:2016oxy,Khanna:2016yow}, where the signal-to-noise ratio is sufficiently large. At such early times, the intrinsic oscillation modes of ECOs, expected to manifest over much longer timescales, have yet to be excited. Consequently, there is still no observational evidence for these long-lived modes, which are theoretically predicted to produce a more complex spectrum compared to that of BHs \cite{Chirenti:2007mk,Pani:2009ss,Maggio:2020jml,McManus:2020lgm,Maggio:2021ans,Ikeda:2021uvc,Saketh:2024ojw,Alfaro:2024tdr,Khoo:2024yeh,Boumaza:2025xgx}. The resulting long-timescale features of the ringdown of an ECO, commonly referred to as GW echoes \cite{Maggio:2017ivp,Maselli:2017tfq,Mark:2017dnq,Maggio:2018ivz,Correia:2018apm,Wang:2018gin,Urbano:2018nrs,Maggio:2019zyv,LongoMicchi:2020cwm,Vlachos:2021weq,Chatzifotis:2021pak,Chatzifotis:2021hpg,Vellucci:2022hpl,Xin:2021zir,Testa:2018bzd,Maggio:2019zyv,Siemonsen:2024snb}, arise from the reflective cavity that forms between the light ring and the ECO's surface. This cavity can repeatedly stimulate emissions reminiscent of the prompt ringdown. While the interpretation of GW echoes as successive excitations of the light ring is physically well-supported, the current QNM framework of ECOs remains incomplete: it does not account for the prompt ringdown\footnote{See though Ref. \cite{DeSimone:2025sgu}.} and is sensitive only to the late-time dynamics of the system \cite{Price:2017cjr}.

Even though the calculation of QNMs of compact objects can be achieved with a variety of methods, and has been streamlined throughout the years \cite{Pani:2013pma}, contemporary phenomena are still arising in BH spectroscopy \cite{Baibhav:2023clw}. Before embarking in new issues of BH spectroscopy let us recall that QNMs do not form a complete basis of eigenfunctions, since they are the solutions of a non-selfadjoint system \cite{Kokkotas:1999bd}. This implies that perturbations to the background BH spacetime cannot be decomposed only in QNMs, since the ringdown displays an early-time onset that is initial-condition-depend, and a late-time Price-law tail \cite{Price:1972pw,Gundlach:1993tp,Barack:1998bw} due to backscattering of perturbations off the BH potential. 
 
Recently, the mathematical tools of \emph{non-modal analysis}, i.e., the study of (non-)selfadjoint operators without the use of eigenvalues, appeared in the field of BH perturbation theory \cite{Jaramillo2020}. It turns out that the non-selfadjoint operator that describes wave propagation in fixed curved spacetimes and, in particular, the QNMs of compact objects are exponentially sensitive to tiny fluctuations to the effective potential; a phenomenon known as \emph{spectral instability} \cite{Aguirregabiria:1996zy,Vishveshwara:1996jgz,Nollert:1996rf,Daghigh:2020jyk,Qian:2020cnz,Jaramillo2020,Jaramillo:2021tmt,Cheung:2021bol,Yang:2024vor,Qian:2024iaq}. The fragility of the BH spectra can be exposed through tiny alterations of the boundary conditions \cite{Leung:1997was,Burenkov:2002,Mirbabayi:2018mdm,Li:2024npg}, by the presence of astrophysical configurations \cite{Cheung:2021bol,Courty:2023rxk,Boyanov:2024fgc,MalatoCorrea:2025iuc,Berti:2022xfj,Spieksma:2024voy} or by near-horizon deviations from the BH paradigm \cite{Cardoso:2016rao,Cardoso:2016oxy,Cardoso:2017cqb,Abedi:2020ujo}. Nevertheless, the prompt ringdown signal is either much less affected by this apparent QNM instability \cite{Berti:2022xfj,Kyutoku:2022gbr,Ianniccari:2024ysv,Shen:2025yiy}, or other quantities indirectly connected to the ringdown appear to be spectrally-stable \cite{Rosato:2024arw,Yang:2022wlm,Volkel:2022ewm,Volkel:2025lhe,Torres:2023nqg,Yang:2024vor,Nair:2025anr}. 

The spectral instability of BH QNMs has been known for at least 30 years \cite{Nollert:1996rf}, but only recently was established as a scientific attribute of QNMs and has been associated with non-modal tools \cite{Jaramillo2020}. Through the apprehension of spectral instabilities, more tools were introduced in the literature that did not base linear stability analyses of compact objects purely on modal decompositions of the ringdown. Non-modal tools, such as the pseudospectrum \cite{Davies:1999,Krejcirik:2014kaa,Jaramillo2020,Destounis:2021lum,Gasperin:2021kfv,Jaramillo:2022kuv,Boyanov:2022ark,Destounis:2023nmb,Sarkar:2023rhp,Boyanov:2023qqf,Arean:2023ejh,Cownden:2023dam,Cao:2024oud,Garcia-Farina:2024pdd,Cai:2025irl,Siqueira:2025lww,dePaula:2025fqt}, pseudoresonances \cite{Reichel:1992,Aslanyan:1998,Trefethen:1993,Driscoll:1996}, and transient effects \cite{Jaramillo:2022kuv,Carballo:2024kbk,Chen:2024mon,Carballo:2025ajx} are currently paving the way towards a clearer understanding of how such effects may affect QNM calculations, as well as how important spectral instabilities can eventually be for BH spectroscopy \cite{Davies:2004,Jaramillo:2021tmt}.

In this work, we focus on spherically-symmetric ECOs that present spectrally-unstable QNMs, with respect to their exterior BH counterpart. We consider an ECO, with the same properties as the one used in Ref. \cite{Boyanov:2022ark}, that has a purely-reflective surface instead of an event horizon, and is parameterized by its compactness, i.e. the proximity of the ECO surface to the Schwarzschild radius. The closer the surface is to the would-be event horizon, the more compact the ECO becomes. We show that environmental effects can further amplify the spectral instability of ECOs, destabilizing overtones and even transforming them into new \emph{``perturbed'' fundamental modes} through an overtaking spectral instability, also found in \cite{Cheung:2021bol}. We further find that the fundamental mode of less-compact exotic objects can be destabilized when the environmental bump moves outward from the light ring. Quite surprisingly, the lowest-lying dominant modes of ultra-compact exotic objects are spectrally-stable, at least in a large physical portion of the parameter space. In finality, due to the robustness of the fundamental mode in ultra-compact ECOs, and the inability of destabilized modes in less-compact ECOs to migrate across the real axis, the overall spectral (in)stability taking place leaves the modal stability of the object intact. 

In what follows we use geometrized units, such that $G=c=1$.

\section{Spherically-symmetric exotic compact objects}

\subsection{Causal structure}

We consider a spherically-symmetric ECO of mass $M$ with a surface $r=r_s$ located at a radial position 
\begin{equation}
\label{eq:ECO_surface}
r_s=r_{h}+\mathcal{E}, \quad \mathcal{E}\ll 1,
\end{equation}
where $r_h=2M$ is the Schwarzschild radius and $\mathcal{E}/r_h$ is the compactness of the ECO. The smaller $\mathcal{E}$ is, the closer the ECO surface is to the Schwarzschild radius and the more compact the ECO becomes. Increasing $\mathcal{E}$ places the ECO surface further away from the Schwarzschild radius, which leads to a less compact ECO.

The geometry outside the ECO is described by the Schwarzschild metric. The line element reads
\begin{equation}\label{Schw}
{\rm d}s^2=-f(r){\rm d}t^2+f(r)^{-1}{\rm d}r^2+r^2{\rm d}\Omega^2_2,
\end{equation}
where $f(r)=1-r_{h}/r$ and ${\rm d}\Omega^2_2$ is the line element of the unit $2$-sphere. For later purposes, we define the tortoise coordinate $r^*$ as
\begin{equation}
\label{eq:tortoise_coord}
r^*=r+r_{h}\log \left(\frac{r}{r_{h}}-1\right),
\end{equation}
where the Schwarzschild event horizon $r\rightarrow r_h$ corresponds to $r^*\rightarrow-\infty$ and infinity $r\rightarrow+\infty$ corresponds to $r^*\rightarrow+\infty$.

\subsection{Scalar perturbations and quasinormal modes}

Perturbation theory in spherically-symmetric backgrounds is typically described by a dynamical field $\Phi(t,r^*)$ obeying the wave equation
\begin{equation}\label{perteq}
\frac{\partial\Phi^2(t,r^*)}{\partial t^2}-\frac{\partial\Phi^2(t,r^*)}{\partial r^*{^2}}+V\Phi(t,r^*)=0,
\end{equation}
where $t$ is a time coordinate parameterizing the evolution, and $r^*$ identifies the spatial domain of integration; $r^*\in (-\infty, +\infty)$ for the standard tortoise coordinate.
The effective potential $V$ depends on the spacetime background and the nature of the perturbation field. Here, we focus on the Regge-Wheeler potential \cite{Berti:2009kk}
\begin{equation}\label{potential}
V=\dfrac{f(r)}{r^2}\left[\ell(\ell+1)+(1-s^2)\frac{r_{h}}{r}\right],
\end{equation}
with $\ell$ the angular number and $s$ the spin parameter characterizing the scalar ($s=0$), vector ($s=\pm1$) and axial tensor ($s=\pm2$) perturbation.

In asymptotically-flat spacetimes, $V(r\rightarrow+\infty)\rightarrow 0$ as $r^*\rightarrow + \infty$. This translates to purely outgoing waves at infinity such that
\begin{equation}
\label{BH_BC_time} 
\frac{\partial\Phi(t,r^*)}{\partial t}+\frac{\partial\Phi(t,r^*)}{\partial r^*}\sim 0,\quad {\rm for} \quad r^* \rightarrow +\infty,
\end{equation}
or, equivalently, in the frequency domain and after a Fourier decomposition of the form $\Phi(t,r^*)\sim \phi(r^*)e^{-i\omega t}$
\begin{eqnarray}
\label{BC_Freq}
\phi(r^*) \sim e^{i\omega r^*} \quad &{\rm for}& \quad r^*\rightarrow +\infty.
\end{eqnarray}
Typically, when an event horizon at $r=r_h$ is present, the potential $V(r\rightarrow r_h)\rightarrow 0$ and the boundary condition is purely ingoing there, i.e.,
\begin{equation}
\label{BH_BC_time2} 
\frac{\partial\Phi(t,r^*)}{\partial t}-\frac{\partial\Phi(t,r^*)}{\partial r^*}\sim 0,\quad {\rm for} \quad r^* \rightarrow - \infty,
\end{equation}
or, in the frequency domain
\begin{eqnarray}
\label{BC_Freq}
\phi(r^*) \sim e^{-i\omega r^*} \quad &{\rm for}& \quad r^*\rightarrow -\infty.
\end{eqnarray}
Modeling ECO boundary conditions, however, needs further consideration. In fact, those can change dramatically at the ECO surface. Similar to neutron stars, we have to have an equation of state of the star's interior that describes how its matter content interacts with perturbations. In these cases, one approach, is to consider a solution in the domain $r^*\in [r^*_s, +\infty)$, where a boundary condition involving a combination of $\Phi(t, r^*_s)$ and its spatial derivative is imposed at the finite point $r^*=r^*_s$ (see e.g. \cite{Maggio:2020jml}). 

Here, we make the assumption that all incoming waves are completely reflected at the ECO's surface, through a Dirichlet boundary condition
\begin{equation}
\label{ECO_BC_time}
\Phi(t,r^*_s)=0.
\end{equation}
One may expect that this is a reasonably good approximation to an ECO which prevents rapid changes in its mass. Moreover, boundary conditions which involve partial absorption and partial reflection would require additional assumptions regarding the nature of the ECO and how its size increases with the absorbed energy \cite{Vellucci2022,Zhong:2022jke}. In our case, we make the simplest assumption in order to study the spectrum and its characteristics on a static ECO with a Dirichlet boundary condition at its surface\footnote{Another boundary condition that can impose pure reflection at the ECO surface is the Neumann boundary $\Phi^\prime(t,r^*)|_{r^*=r^*_s}=0$, where a prime denotes differentiation with respect to the tortoise radius. We expect the qualitative behavior of the resulting QNMs to be similar, as in Refs. \cite{Cardoso:2016rao,Cardoso:2016oxy}.}. 

In principle, a horizonless compact object with pure reflectivity surrounded by a matter halo (that introduces the ``flea''), could be constructed with the Einstein cluster formalism. For example, in the supplemental material of Ref. \cite{Cheung:2021bol}, where a procedure is described that embeds a non-thin matter halo to any spherically-symmetric spacetime is presented. What would change in our current object is its reflectivity at its surface. To surpass this issue we could utilize a spherically-symmetric ansatz, such as that of a gravastar or a thin matter shell \cite{Cardoso:2016oxy}, which have well-defined boundary conditions at their surfaces. In fact, similar constructions have been found, for example, in \cite{Konoplya:2018yrp} where an exotic compact object is constructed to have a matter shell around it, in order to understand how echoes of the surface of the ECO are affected by the surrounding astrophysical environment. Also, since it is now widely established that globular clusters host populations of white dwarf, neutron star, and BH binaries, dynamical processes enabled by stellar densities millions of times larger than typical galactic environments, facilitate interactions involving these stellar remnants that give rise to an array of astrophysical phenomena \cite{Kremer:2025mae}. Nevertheless, for a full understanding of astrophysical environments and their interaction with ultra-compact objects, the development of a solid theoretical framework should be developed. Since a proper modeling of the coalescence, ringing and data analysis of such objects is mostly lacking, we use the simplified analysis above to gain an initial insight.

To study the spectral content of these objects, we use Eq. \eqref{perteq}. After a Fourier decomposition of the form $\Phi(t,r^*)\sim \phi(r^*)e^{-i\omega t}$, we obtain the frequency-domain master equation
\begin{equation}\label{pertF}
\frac{d^{2}\phi(r^*)}{dr^*{^2}}+\left(\omega^2-V\right)\phi(r^*)=0.
\end{equation}
The QNMs of the ECO are solutions of Eq.~\eqref{pertF}, $\omega_n$, where $n=0,\,1,\,2\dots$ is the overtone index, that satisfy a purely outgoing boundary condition at infinity and an appropriately reflective boundary condition at the ECO's surface. In particular, the Fourier space equivalent to the boundary conditions \eqref{BH_BC_time} at infinity, and to \eqref{ECO_BC_time} at the surface, reads
\begin{align}
\label{BC_Freq}
\phi(r^*) &\sim e^{i\omega r^*}, \quad \quad r^*\rightarrow +\infty, \\ \label{BC_Freq_surf}
  \phi(r^*_s) &= 0,\qquad\quad\,\,\,\,r^*=r^*_s.
\end{align}

\subsection{Including environments}

To include an astrophysical (or otherwise) environments around a compact object such as a BH or an ECO, one can utilize the Einstein cluster formalism which chooses a stress-energy tensor that describes the environment as an anisotropic fluid with specific energy density and tangential pressure. This method has been widely used recently and has led to new general-relativistic solutions that describe static, spherically-symmetric, as well as axially-symmetric rotating BHs in dark matter halos \cite{Cardoso:2021wlq,Fernandes:2025osu}.

\begin{figure}[t]
    \centering
    \includegraphics[width=\linewidth]{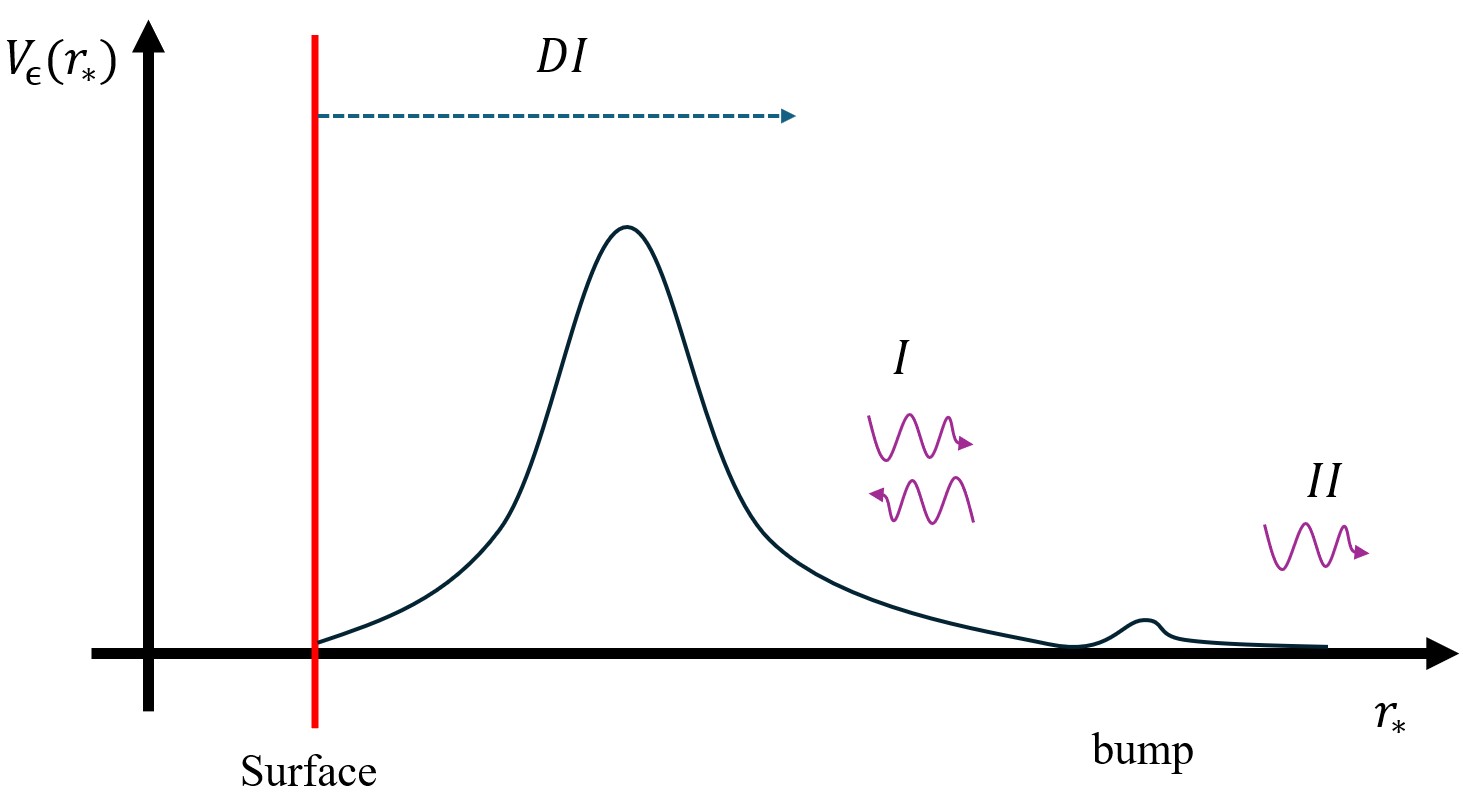}
    \caption{Illustrative picture of the integration procedure. We start at the surface of the compact object, represented by the vertical solid red line, using the imposed boundary conditions, directly integrating to the region $I$ (DI path). At the region $I$, the solution is matched with two continued fraction expansions that carry the information of ingoing and outgoing wave behavior. Finally, we integrate again through the bump, matching the solution in the region $II$ with a purely outgoing one.}
    \label{fig:schematics}
\end{figure}

The Einstein formalism can also be utilized in order to, not only include matter halos around compact objects, but also to embed thinner halos of matter around them (see Fig. 3 in the Supplemental material of Ref. \cite{Cheung:2021bol}). The resulting effective potential of such general-relativistic solution has a qualitatively similar form as the one shown in Fig. \ref{fig:schematics}, i.e. a main potential peak that corresponds to the light ring of the ECO (\emph{the ``exotic'' elephant}) and an orders-of-magnitude smaller bump (\emph{the flea}) in some distance from the main potential (not shown in scale in Fig. \ref{fig:schematics}). Therefore, we can simplify the analysis and instead of building a general-relativistic solution of an ECO with a thin matter halo, we can add the bump by hand in the effective potential of the ECO described by Eq. \eqref{potential}. The resulting spectra behave in a similar qualitative way, as shown in \cite{Cheung:2021bol}.

We will use the perturbed potential 
\begin{equation}\label{eq:perturbed_potential}
    V_\epsilon=V+\epsilon\, V_\textrm{bump},
\end{equation}
where $V$ is the Schwarzschild potential outside of the ECO and $\epsilon$ is the scale of the bump. We choose a Gaussian distribution to mimic the astrophysical bump outside of the ECO, that is
\begin{equation}\label{eq:Gaussian_bump}
    V_\textrm{bump}(r^*-a_0)=\exp\left( -\frac{(r^*-a_0)^2}{2\varrho}\right),
\end{equation}
where $r^*=a_0$ is the mean of the distribution, i.e. the radius that the Gaussian peak is centralized, and $\varrho$ is the standard deviation of the distribution.

By substituting the potential $V$ with the perturbed one from Eq. \eqref{eq:perturbed_potential} in the master equation \eqref{pertF} we can apply the appropriate boundary conditions, previously discussed, for an ECO with a perfectly reflecting surface in order to obtain the new \emph{perturbed QNMs}, due to the slight modification introduced through the Gaussian distribution bump.

\section{Numerical methods}

In what follows, we describe the two main numerical techniques that we used to calculate both the vacuum and perturbed ECO QNMs, i.e., the hyperboloidal scheme \cite{Ansorg2016,PanossoMacedo:2018hab,PanossoMacedo:2020biw,PanossoMacedo:2023qzp,PanossoMacedo:2024nkw} and the continued fraction method \cite{Leaver1985,Onozawa:1995vu,Richartz:2015saa}. All results shown below are those obtained from the hyperboloidal scheme. A complete discussion of  further QNM techniques can be found in Refs. \cite{Berti:2009kk,Kokkotas:1999bd,Konoplya:2011qq,Pani:2013pma}. 

\subsection{Hyperboloidal approach to ECO QNMs}

\begin{figure*}[t]
    \centering
    \includegraphics[width=\linewidth]{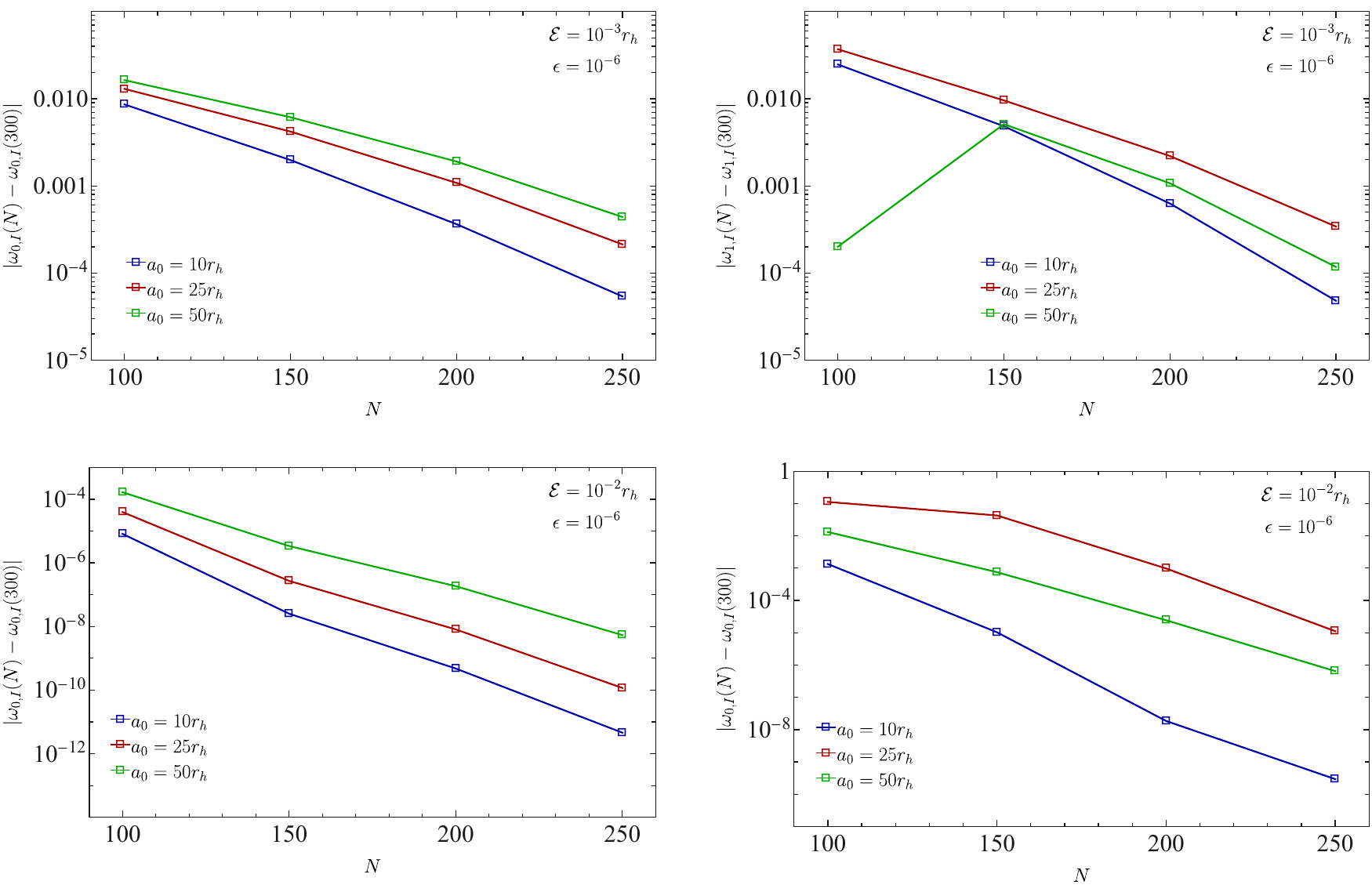}
    \caption{\emph{Top row:} Convergence tests of the fundamental $\ell=2$ mode ($n=0$, top left) and first overtone ($n=1$, top right) for an ECO with $\mathcal{E}/r_h=10^{-3}$ and a bump at varying positions $a_0$ and amplitude $\epsilon=10^{-6}$. Due to the absence of of an exact spectrum, we have used the $N=300$-grid-points value of QNMs as our baseline and compared its convergence by varying $N$. We find that the absolute difference of the imaginary parts between the $N=300$ and, say, $N=250$ is of order $\mathcal{O}(10^{-3}-10^{-4})$. The same order of magnitudes are found for the real part of the QNMs compared. \emph{Bottom row:} Same as above but for an ECO with $\mathcal{E}/r_h=10^{-2}$. We find that the absolute difference of the imaginary parts between the $N=300$ and, say, $N=250$ is of order $\mathcal{O}(10^{-8}-10^{-10})$ for the fundamental mode (bottom left) and $\mathcal{O}(10^{-4}-10^{-8})$. The same order of magnitudes are found for the real part of the QNMs compared. The exponential convergence in all plots confirms the expected behavior from the employed spectral scheme.}
    \label{fig:convergence}
\end{figure*}

To obtain the QNMs, we make use of the hyperboloidal framework via the scri-fixing scheme combined with a spatial compactification \cite{Zenginoglu:2007jw,Zenginoglu:2011jz,PanossoMacedo:2023qzp, PanossoMacedo:2024nkw}. We perform a change of coordinate from $(t,r)$ to $(\tau,\sigma)$, given by
\begin{equation}
    t=\tau-H(\sigma), \hspace{0.5 cm} r=\frac{r_{h}}{\sigma}.
\end{equation}
In the case of BH spacetimes, we have $\sigma \in [0,1]$, with the event horizon and null infinity at $\sigma=1$ and $\sigma=0$, respectively. These coordinates naturally incorporate the outgoing (null infinity) and the ingoing (event horizon) behavior of the field as given by Eqs.~(\ref{BH_BC_time}) and (\ref{BH_BC_time2}). Using the Fourier decomposition as $\Phi=e^{-i\omega \tau}\psi(\sigma)$, we obtain the differential equation
\begin{equation}\label{EQ:waveEqSigma}
        \begin{aligned}
        \frac{d}{d\sigma}\left(p(\sigma)\frac{d\psi(\sigma)}{d\sigma}\right)+&\left(\omega^{2}w(\sigma)-\frac{V(\sigma)}{p(\sigma)}\right)\psi(\sigma)+ \\
        -&i\omega \left(2\gamma(\sigma)\frac{d\psi(\sigma)}{d\sigma}+\gamma'(\sigma)\psi(\sigma)\right)=0,
        \end{aligned}
\end{equation}  
where
\begin{equation}
    \begin{aligned}
        p(\sigma)=-\frac{1}{dr^{*}(\sigma)/d\sigma},\\
        \hspace{0.2 cm} \gamma(\sigma)=\frac{dH}{d\sigma}p(\sigma), \\
        w(\sigma)=\frac{1-\gamma(\sigma)^2}{p(\sigma)}.
    \end{aligned}
\end{equation}
In this work, we will analyze ECOs with purely reflective surfaces, i.e., apply the boundary condition \eqref{ECO_BC_time}. The hyperboloidal framework presents the same equations, but the spacetime is mapped to the interval $\sigma \in [0,\sigma_{0}]$, with $\sigma_{0}=r_{h}/(r_{h}+\mathcal{E})<1$, where at $\sigma_{0}$ we impose the Dirichlet boundary condition. This was employed in Ref. \cite{Boyanov:2022ark} for calculating the pseudospectra of ECOs.
    
We perform a first-order reduction in time $\bar{\phi}=d\psi/d\tau=-i \omega\psi$, that leads to the system of equations
\begin{equation}	
    \mathbf{L}\vec{U} =-i\omega \vec{U},
\end{equation}
with
\begin{equation}\label{Eq:L_and_U}
    \mathbf{L}=\left(\begin{matrix}
    0 & 1 \\
    w(\sigma)^{-1}L_{1} &	\, w(\sigma)^{-1}L_{2}\\
    \end{matrix}\right),\hspace{0.5 cm}
        \vec{U}= \left(\begin{matrix}
		\psi  \\
		\bar{\phi}\\
    \end{matrix}\right),
\end{equation}
The components of the $\mathbf{L}$ operator are given by
\begin{equation}
    \begin{aligned}
        L_{1}=\frac{d}{d\sigma}\left(p(\sigma)\frac{d}{d\sigma}\right)-\frac{V(\sigma)}{p(\sigma)},\\
        L_{2}=2 \gamma(\sigma)\frac{d}{d\sigma}+\gamma'(\sigma).
    \end{aligned}
\end{equation}   
To find the eigenvalues of the operator $\mathbf{L}$, we use a pseudospectral method, that expands the eigenfunctions in a set of basis functions as
\begin{equation}
    \psi(\sigma)=\sum_{i=0}^{N}c_{i}T_{i}(\sigma),
\end{equation}
where we use the Chebyshev polynomials, given by $T_{j}(\xi)=\cos(j\,\arccos \,\xi)$. We use the Chebyshev-Lobatto grid for the collocation points and find a matrix of dimensions $(N+1)\times(N+1)$. The Dirichlet boundary condition is imposed by identifying the corresponding row and column of the matrix that corresponds to the ECO surface and setting them to zero. Hence, both boundary conditions \eqref{BC_Freq} and \eqref{BC_Freq_surf} are satisfied \cite{Boyanov:2022ark}. After the boundary conditions are imposed we find the QNM frequencies as the system's eigenvalues. 

Adding the bump to the problem increases the error when approximating the functions using the Chebyshev polynomials. To overcome this issue, we use a double domain approach, where we divide the spacetime into two regions:  (i) between the ECO and the bump, which in the hyperboloidal coordinates is $\sigma \in \sigma_{1}=[\sigma(a_{0}),\sigma_{0}]$, and (ii) between the bump and null infinity, i.e., $\sigma \in \sigma_{2}=[0,\sigma(a_0)]$. Then, we proceed with the Chebyshev polynomials in a Chebyshev-Lobatto grid for each domain, that lead to two operators $\mathbf{L}_{\sigma_{1}}$ and $\mathbf{L}_{\sigma_{2}}$, with the same shape and dimension of $\mathbf{L}$ in Eq.~(\ref{Eq:L_and_U}). Since the spacetime is divided in two sectors, we must impose regularity of the wavefunctions and their derivatives at the location of division, i.e., the bump. By doing so, we reach a generalized eigenvalue problem given by
\begin{equation}	
    \mathbf{L}_{T}\vec{U}_{T} =-i\omega \mathbf{B}\vec{U}_{T},
\end{equation}
with
\begin{equation}
    \mathbf{L}_{T}=\left(\begin{matrix}
    \mathbf{L}_{\sigma_{1}} & \mathbf{BC}_{1}\\
    \mathbf{BC}_{2}&	\mathbf{L}_{\sigma_{2}}\\
    \end{matrix}\right),\hspace{0.5 cm}
        \vec{U}_{T}= \left(\begin{matrix}
		\psi_{\sigma_{1}}  \\
        \phi_{\sigma_{1}} \\
		\psi_{\sigma_{2}}\\
        \phi_{\sigma_{2}}\\
    \end{matrix}\right).
\end{equation}
The final matrix, $\mathbf{L}_{T}$, has dimensions of $2(N+1) \times 2(N+1)$. The transition conditions are imposed by modifying two rows in $\mathbf{L}_{T}$, which results in a single row with at least one non-zero value in the $\mathbf{BC}_{1,2}$ matrices, and in $\mathbf{B}$ which is an identity matrix, besides the two corresponding rows which must be zero. These transition conditions impose the continuity of the function and its derivative at the bump location, and the Dirichlet condition imposes the ECO condition, such that solving this generalized eigenvalue problem, we find the QNMs of the system. The data generated and presented in the rest of this work have been obtain with $N=300$ collocation points to ensure numerical convergence, as shown in Fig. \ref{fig:convergence}. In fact, it is noteworthy to mention that the convergence increases rapidly when the ECO is less compact.
        
\subsection{Continued fraction}

The continued fraction (CF) method is devised to prescribe an \textit{analytical} prescription for the wave function in a vacuum region \cite{Leaver1985,Onozawa:1995vu,Richartz:2015saa}. The expression is chosen such that the outer boundary condition (outgoing waves) is satisfied. Then, we tie the wave function by performing a shooting for the frequency such that the boundary conditions at the ECO and the vacuum region are satisfied. One way to do this is to integrate the wave equation directly from the surface, over the potential bump, and match the solution to a CF expression. This has been explored in Ref.~\cite{MalatoCorrea:2025iuc}. Here we exploit a different integration scheme. This is necessary to cover cases where the bump is located far from the ECO, where the direct-integrated solution gets contaminated from the exponential behavior of QNM solutions.

Firstly, we need to impose the boundary condition at the ECO surface. This is done by setting $\phi(r_s)=0$ and $\phi'(r_s)=1$. We then integrate the wave equation from the surface up to a point where $r_{I}\gtrsim3M$, in which the light-ring potential peak is practically negligible, i.e., $V\sim\delta V$. The schematic is illustrated in Fig.~\ref{fig:schematics}. This allow us to construct the directly-integrated (DI) solution $\phi_{\rm DI}(r)$ that satisfies the ECO boundary condition.

At the region $I$ depicted in Fig.~\ref{fig:schematics}, the generic behavior is given by a combination of ingoing and outgoing wave solutions. Each of those can be analytically constructed by a series expansion. For example, the outgoing solution is obtained through
\begin{equation}\label{eq:series}
    \phi(r)=\chi(r)\sum_{n=0}^N b_n v(r)^n,
\end{equation}
where $v(r)=1-d/r$, $\chi=e^{i\omega r}(r/r_h-1)^{ir_h\omega}$, and $N$ is some large number at which the series is truncated. The constant $d$ is chosen such that the above series is convergent \cite{Benhar:1998au}. By replacing the above into the field equation \eqref{pertF}, we obtain a four-term recurrence relation for the coefficients $b_n$, such that
\begin{equation}
    \alpha_n b_{n+1}+\beta_n b_n+\gamma_nb_{n-1}+\delta_n b_{n-2}=0,
\end{equation}
where
\begin{align}
    \alpha_n&=[(d - 2 M) n (1 + n)]/d,\\
    \beta_n&=2 n \left[n \left(\frac{3 M}{d}-1\right)+i d \omega \right],\\
    \gamma_n&=\frac{2 M \left[-3 (n-1) n+s^2-1\right]}{d}\nonumber\\
    &-(\ell-n+1) (\ell+n),\\
    \delta_n&=\frac{2 M \left[(n-1)^2-s^2\right]}{d}.
\end{align}
The above relation can be solved by standard methods, giving all the coefficients as functions of $a_0$ (see, e.g., \cite{Pani:2013pma}). By setting $a_0=1$ we have constructed a series solution that is purely outgoing, namely $\phi_+(r)$, given by the analytical series \eqref{eq:series}. It is straightforward to show that an ingoing solution, say $\phi_-(r)$, can be obtained by simply performing the transformation $\omega\to-\omega$. The general solution in region $I$ can be written as a linear combination of the ingoing and outgoing solutions, i.e.,
\begin{equation}
    \phi_I(r)=c_1 \phi_+(r)+c_2 \phi_-(r),
\end{equation}
where $c_1$ and $c_2$ are obtained by imposing continuity of $\phi_I$ and its first derivative at a point $r_I$ with the solution integrated from the ECO surface $\phi_{\rm DI}$.

It is worth to note that from the above prescription, the standard QNM problem without the environmental bump requires that $c_2=0$, which is the condition that selects a set of complex frequencies $\omega_{n}$. We have verified that this is indeed the case, recovering the usual modes of relativistic stars \cite{Redondo-Yuste:2025ktt}. When the potential includes a small bump, however, an additional step is required.

We start at the region $I$ (see Fig.~\ref{fig:schematics}), from the left of the bump at a point where $\delta V$ is negligible compared to $\epsilon$, i.e. the perturbed potential satisfies $V_\epsilon\sim\epsilon V_\textrm{bump}$. As boundary conditions, we use the solution $\phi_I$ previously integrated. We then proceed to integrate over the bump, to the region $II$, again to a point where $\delta V$ is negligible. Finally, we demand that this solution, i.e. $\phi_{II}$, is linearly dependent to a purely outgoing solution. This solution obeys the same recurrence relations shown previously, though a different value for the constant $d$ should be chosen to ensure that the series is convergent. Therefore, the QNMs are found by requiring that the Wronskian is zero, i.e.,
\begin{equation}
    W(\phi_+,\phi_{II})=0.
\end{equation}

With the above procedure, we are able to calculate the fundamental mode and the first few overtones, depending on the location of the bump. The agreement with the hyperboloidal approach described previously is excellent, and is illustrated in Table \ref{tab:mode_comp}.

\section{The role of the ECO compactness on QNMs}

\begin{figure}
    \centering
    \includegraphics[width=\columnwidth]{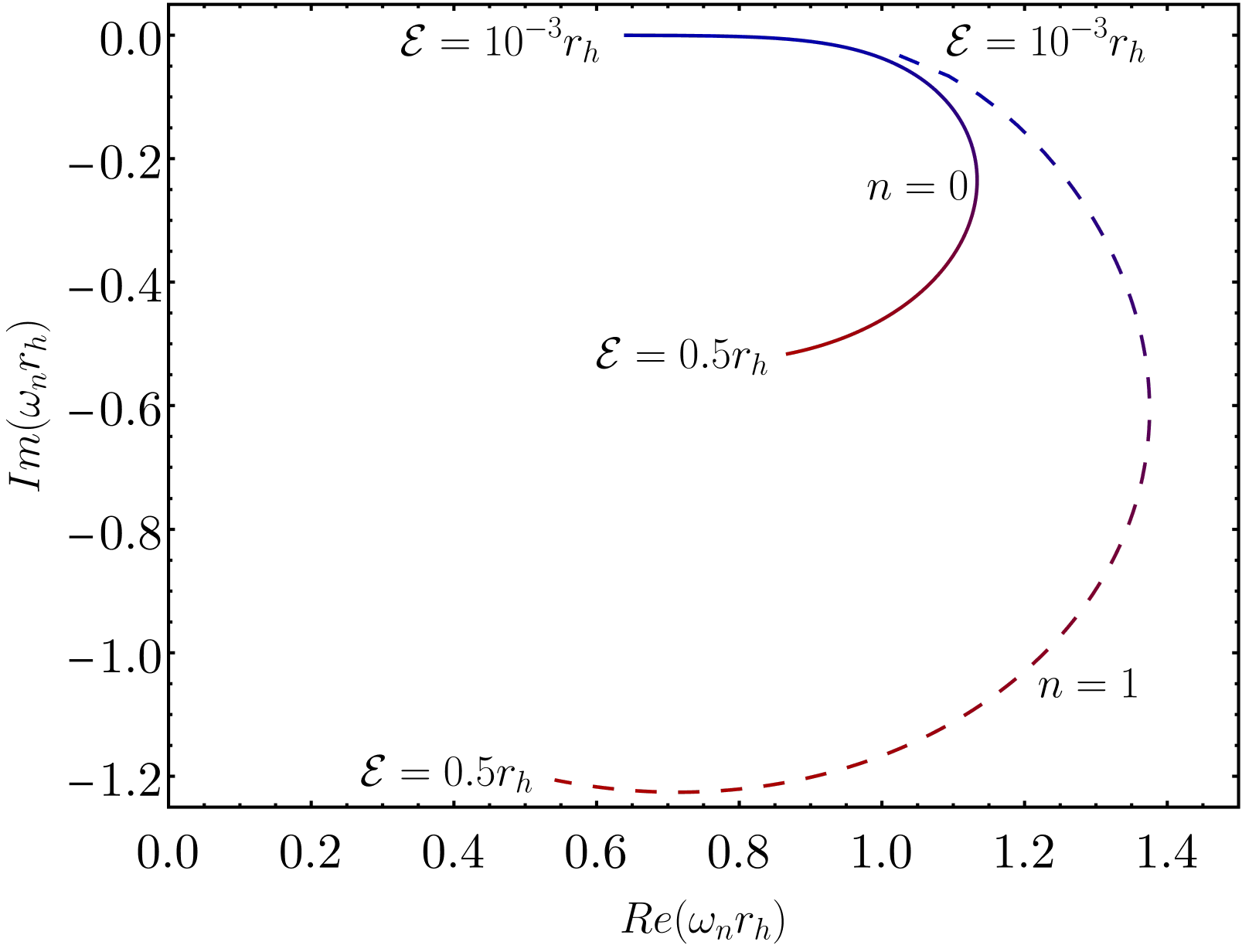}
    \caption{Scalar massless $\ell=2$ QNMs of an ECO, with a perfectly reflective surface at $r=r_s$. By varying the surface position from $\mathcal{E}/r_h=10^{-3}$ to $0.5$ we plot the evolution of the first two modes, i.e. the fundamental $n=0$ and first overtone $n=1$, in the complex plane. The change of the color from blue to red designates the increment of $\mathcal{E}/r_h$.}
    \label{fig:ECO_QNMs}
\end{figure}

The modes discussed here serve as a validity test for our codes and how well they behave with varying ECO compactness $\mathcal{E}/r_h$. The QNM data are filtered by searching for modes from both codes that agree up to precision of order $\mathcal{O}(10^{-6})$. In what follows, we set $r_h=2M=1$.

\begin{table*}
    \centering
    \begin{tabular}{c|c|c|c}
    \hline\hline
    $a_0$ & $n$ & Hyperboloidal scheme &  Leaver's method \\\hline
    {$10 r_h$} & \begin{tabular}{@{}c@{}} 0\\1 \end{tabular} & \begin{tabular}{@{}c@{}} $1.1334624190701 -0.234523615651421i$ \\ $0.8103841283438 - 0.635580461869585i$ \end{tabular} & \begin{tabular}{@{}c@{}}
     $1.13346241914305 - 0.234523615287228i$ \\ $0.81038412927508 - 0.635580462800902i$ \end{tabular} \\
      \hline
         {$25 r_h$} &\begin{tabular}{@{}c@{}} 0\\1 \end{tabular} & \begin{tabular}{@{}c@{}} $1.1377111041509-0.228417661482553i$\\ $1.0688001955232-0.263400692101339i$ \end{tabular}
                   & \begin{tabular}{@{}c@{}} $1.13771110412531 - 0.228417661639658i$ \\ $1.06879997383548 - 0.263400586221004 i$ \end{tabular} \\
      \hline
         {$50 r_h$} &\begin{tabular}{@{}c@{}} 0\\1 \end{tabular} & \begin{tabular}{@{}c@{}} $0.1695236173071-0.127254357355847i$ \\ $0.2354620386002-0.127745914123775i$ \end{tabular} & \begin{tabular}{@{}c@{}}
                 $0.16952361732946 - 0.127254357344676i$ \\ $0.23546203860936- 0.127745914099124i$ \end{tabular} \\
      \hline\hline
    \end{tabular}
    \caption{Scalar $\ell=2$ QNM frequencies computed through the hyperboloidal scheme and the continued fraction (Leaver's) method. We consider an ECO with a Gaussian bump, with $\epsilon=10^{-6}$ and $\varrho=1$, at locations $a_{0}/r_h=10,\,25$ and $50$. The ECO compactness is chosen as $\mathcal{E}/r_h=10^{-1}$. We present the fundamental mode ($n=0$) and the first overtone ($n=1$) for all cases. The number of significant figures are chosen to illustrate the difference between the methods, although it does not necessarily expose the precision of the computation.
    }
    \label{tab:mode_comp}
\end{table*}

The behavior of the spectrum under changes of the boundary conditions has been discussed extensively in Refs. \cite{Cardoso:2016rao,Cardoso:2016oxy}. The mechanism behind the existence of long-lived QNMs depends on the trapping region that forms between the ECO's reflective surface and the light ring. The closer the ECO surface is placed to the Schwarzschild radius, the more effective the interior trapping region due to its enlargement. The ECO's spectrum is therefore an outcome of the \emph{interior trapped modes} that excite the object. We vary the compactness smoothly in order to understand if the ECO spectrum will vary in a similar manner; an assumption that we confirm in Fig. \ref{fig:ECO_QNMs}. We slowly increase the ECO surface location, where the maximum $\mathcal{E}/r_h=0.5$ places the surface close to the light ring. Mode migration in the complex plane is proportional to the order-of-magnitude change of $\mathcal{E}/r_h$. The compactness of ECOs does not affect the spectrum in a discontinuous manner, even though the parameter $\mathcal{E}$ effectively changes the position of the boundary condition. We note that the formation of the stable trapping regime is also the reason behind the existence of late-time GW echoes in the ringdown signal of ECOs \cite{Cardoso:2016rao,Cardoso:2016oxy,Vlachos:2021weq,Chatzifotis:2021pak,Ikeda:2021uvc,Cardoso:2019apo,Coates:2019bun,Coates:2021dlg,Agullo:2020hxe,Cardoso:2019rvt,LongoMicchi:2020cwm,Urbano:2018nrs,Xin:2021zir}.

Interpreting the absence of a spectral instability with respect to the compactness of the ECO can be realized in two ways: (i) the ECO acts as a typical object where its dynamics continuously change when we vary the available parameters. Note though that the surface position is added \emph{by hand} and does not backreact to the spacetime metric, which is otherwise Schwarzschild besides the boundary condition at the ECO surface. (ii) In a more theoretical and counter-intuitive level of understanding spectral instabilities, and their connection to the pseudospectrum, shown in Ref. \cite{Boyanov:2022ark}, we are utilizing an object that has already undergone a drastic change of features with respect to its exterior BH counterpart. In other words, contrary to the QNM spectral instability paradigm, where small perturbations in the system lead to a large deviation in the spectrum, the ECO modes can be considered as a dramatically destabilized spectral set with respect to their BH QNM counterpart, because a very large modification has been introduced to the system, in the form of a complete alteration of the near-horizon boundary condition. However, once the new QNMs arise, the spectra is stable with respect to small deviation of the parameter controlling the location of the boundary condition.

Interestingly, previous studies have shown that when the effective potential of a Schwarzschild BH is perturbed, the resulting QNM spectrum can migrate significantly in the complex plane, i.e., a spectral instability occurs. In particular, Ref. \cite{Jaramillo2020} reports that discontinuous modifications of the system ---modeled there by random perturbations of the potential--- tend to maximize the QNM migration into regions determined by the pseudospectrum. Remarkably, the addition of further perturbations to such an already perturbed system does not induce additional instabilities. Instead, the pseudospectrum of the perturbed operator develops features typical of spectrally-stable spectra (concentric pseudospectral contours surrounding the new ``perturbed'' eigenvalues), which is interpreted as an improvement in the analytic behavior of the Green’s function (or the resolvent operator). This phenomenon raises a natural question: can a spectrum that has already been destabilized be further destabilized? In Ref. \cite{Siqueira:2025lww} this scenario is explored in detail, demonstrating that destabilized QNMs originating from physically-motivated deformations of the spacetime remain susceptible to further spectral instability. Furthermore, when multiple perturbation sources are present in the system, identifying the origin of the instability may become extremely intricate.

Our analysis in this work may be viewed in the same light: the ECO spectrum can be regarded as the spectrally-unstable QNM counterpart of a Schwarzschild BH. Analogously to the discontinuous character of random perturbations added to the potential, the drastic change in the object’s boundary condition also introduces a discontinuous modification of the original Schwarzschild operator. The absence of a second instability when varying the ECO surface position might therefore reflect not a new stabilizing mechanism but simply another spacetime parameter shifting a boundary condition. However, as we will show below, the addition of a matter halo around the ECO can indeed trigger further spectral instability, providing a physically-motivated example of the scenario envisaged in Ref. \cite{Jaramillo2020} and further explored by Ref. \cite{Siqueira:2025lww}.

\begin{figure*}
    \centering
    \includegraphics[width=\textwidth]{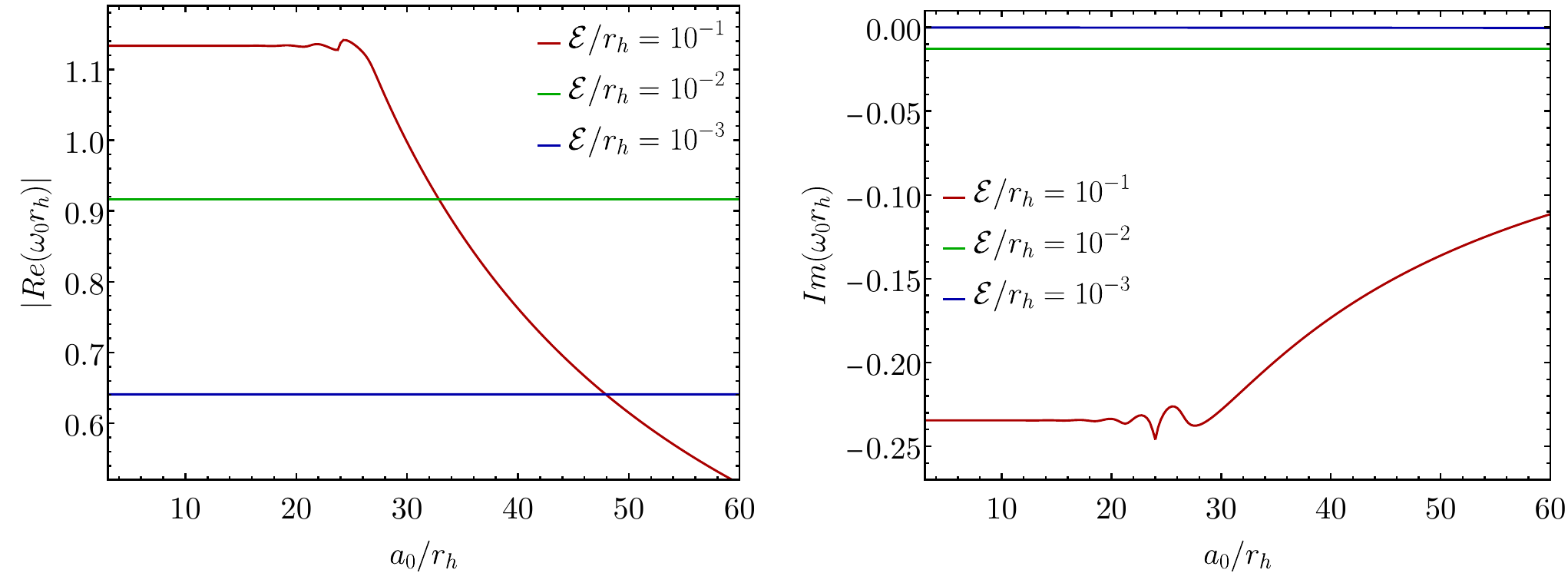}
    \caption{Scalar massless fundamental ($n=0$) $\ell=2$ QNMs of an ECO, with a perfectly reflective surface at $r=r_s$, and a Gaussian bump with $\epsilon=10^{-6}$ and $\varrho=1$ center at $a_0$. By varying the position of the bump from $a_0/r_h=2$ to $60$ we plot the real (left panel) and imaginary part (right panel) of the fundamental mode, for three different compactness, i.e. $\mathcal{E}/r_h=10^{-1}$ (red), $\mathcal{E}/r_h=10^{-2}$ (green) and $\mathcal{E}/r_h=10^{-3}$ (blue).}
    \label{fig:ECO+bump_fund_QNMs}
\end{figure*}

\begin{figure*}
    \centering
    \includegraphics[width=\textwidth]{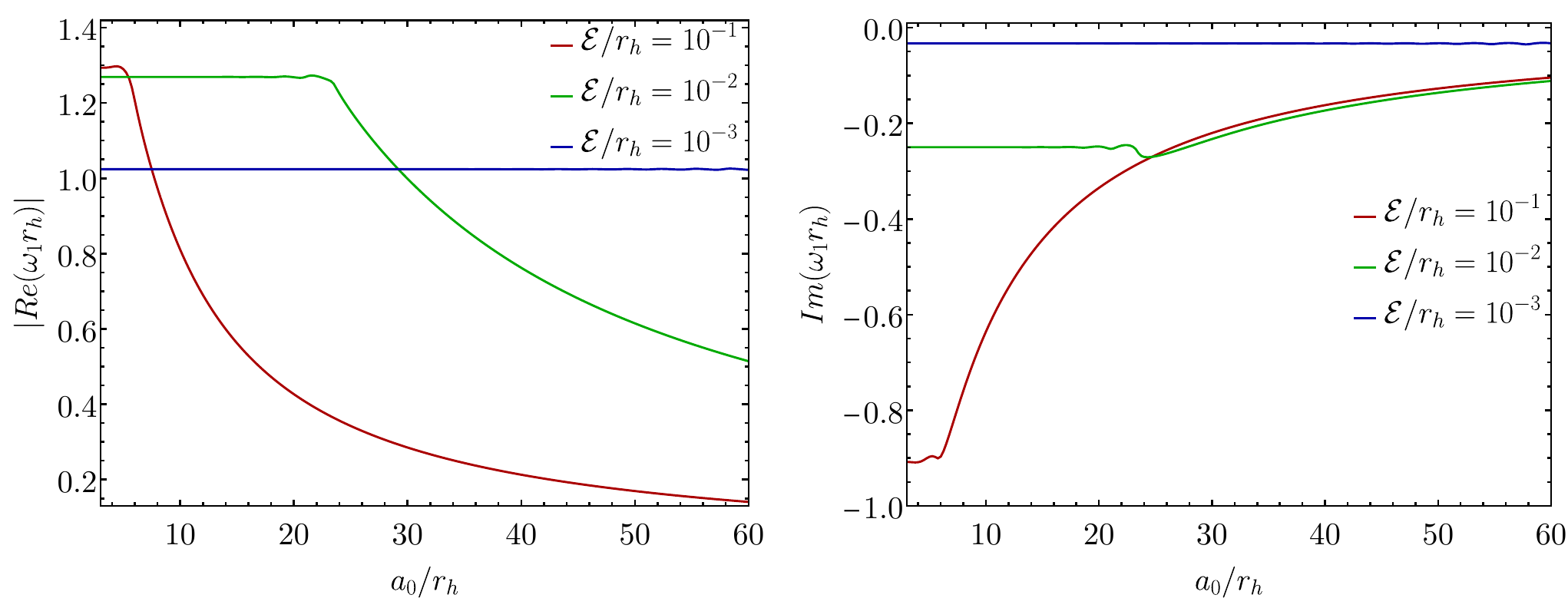}
    \caption{Scalar massless first overtone ($n=1$) $\ell=2$ QNMs of an ECO, with a perfectly reflective surface at $r=r_s$, and a Gaussian bump with $\epsilon=10^{-6}$ and $\varrho=1$ center at $a_0$. By varying the position of the bump from $a_0/r_h=2$ to $60$ we plot the real (left panel) and imaginary part (right panel) of the fundamental mode, for three different compactness, i.e. $\mathcal{E}/r_h=10^{-1}$ (red), $\mathcal{E}/r_h=10^{-2}$ (green) and $\mathcal{E}/r_h=10^{-3}$ (blue).}
    \label{fig:ECO+bump_over_QNMs}
\end{figure*}

\begin{figure*}
    \centering
    \includegraphics[width=\textwidth]{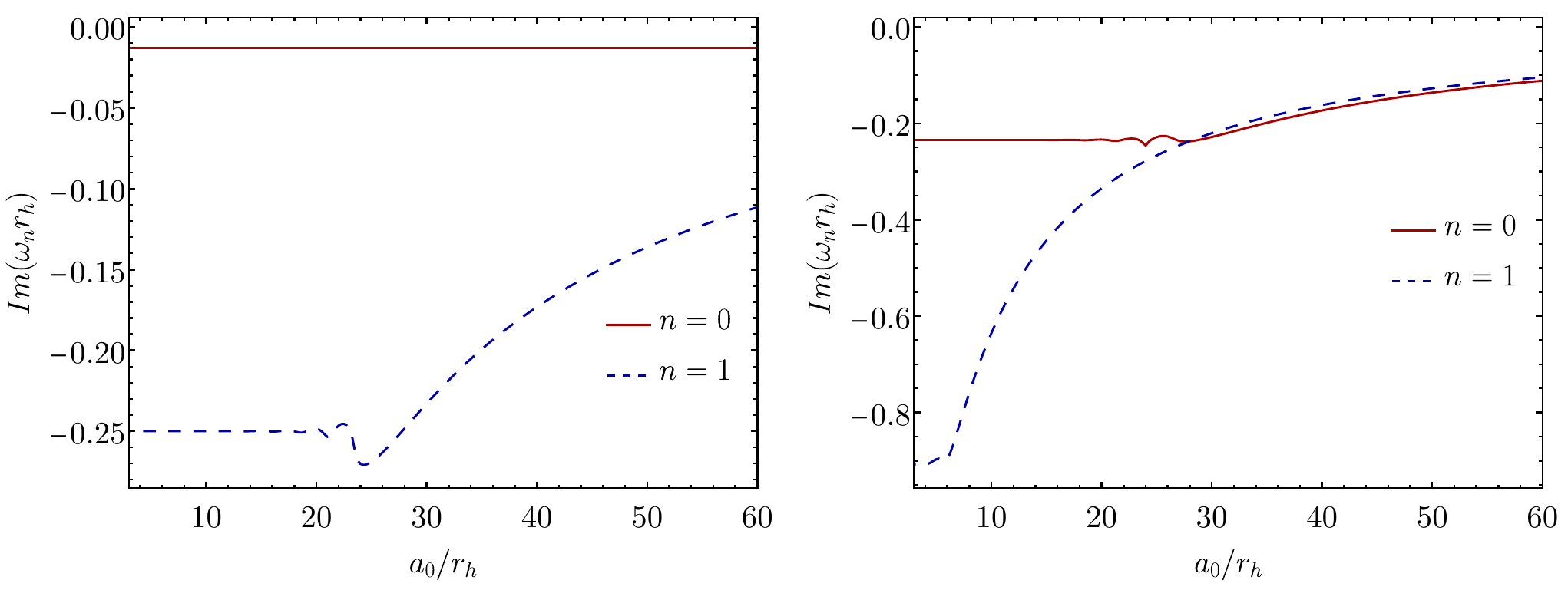}
    \caption{Scalar massless $\ell=2$ QNMs of an ECO, with a perfectly reflective surface at $r=r_s$, and a Gaussian bump with $\epsilon=10^{-6}$ and $\varrho=1$ center at $a_0$. By varying the position of the bump from $a_0/r_h=2$ to $60$ we plot the imaginary parts of the fundamental mode and first overtone for an ECO with compactness $\mathcal{E}/r_h=10^{-2}$ (left panel) and $\mathcal{E}/r_h=10^{-1}$ (right panel).}
    \label{fig:ECO+bump_overtaking}
\end{figure*}

\section{Environmental effects on perturbed ECO QNMs}

In this section, we introduce a localized bump in the effective potential of horizonless ECOs, that mimics environmental effects. By fixing the compactness of the ECO, and varying the position of the localized bump $r^*=a_0/r_h$, described by Eqs. \eqref{eq:perturbed_potential} and \eqref{eq:Gaussian_bump}, we calculate the resulting lowest-lying modes during their dynamical migration in the complex plane. For scalar perturbations on the ECO background, we choose $\ell=2$, while the Gaussian bump has fixed amplitude $\epsilon=10^{-6}$ and standard deviation $\varrho=1$.

\subsection{Destabilization of ECO QNMs}

\begin{figure*}
    \centering\hspace{-1.5cm}
    \includegraphics[scale=0.7]{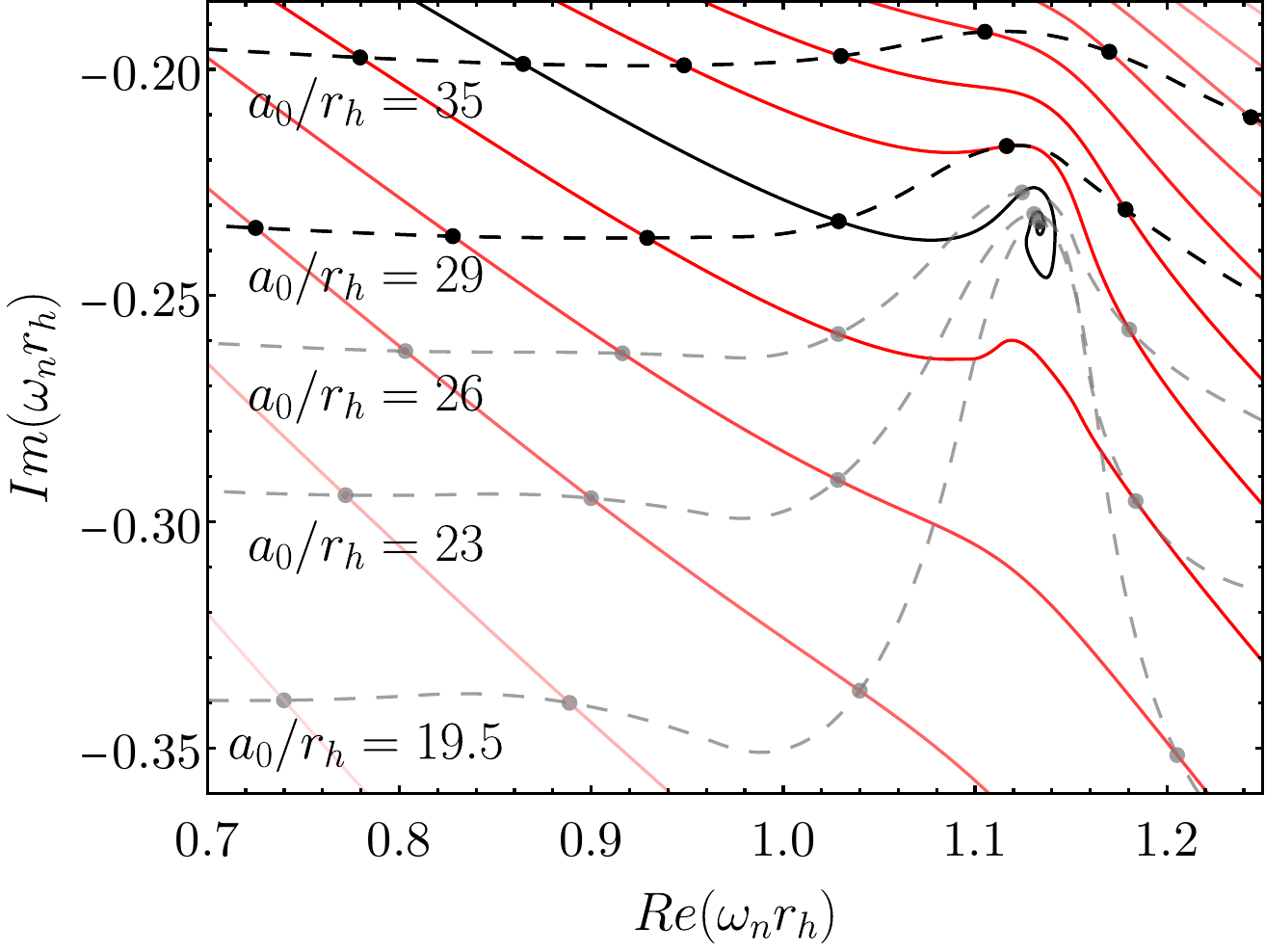}
    \caption{Scalar massless $\ell=2$ QNMs of an ECO ($\mathcal{E}/r_h=10^{-1}$), with a perfectly reflective surface at $r=r_s$, and a Gaussian bump with $\epsilon=10^{-6}$ and $\varrho=1$ center at $a_0$. By varying the position of the bump from $a_0/r_h=2$ to $60$ we plot the QNMs and their migration in the complex plane. The black outspiral curve corresponds to the migration of the ECO fundamental mode, while the red curves (with varying opacity) designates the modes resulting from stable trapping between the light ring and the bump. The dashed gray horizontal-like curves correspond to the QNMs (both ECO and outer bump ones, shown in gray dots) for fixed $a_0/r_h$. In these cases, the ECO QNMs dominate the late-time signal. The dashed black horizontal-like curves correspond to the QNMs for a fixed $a_0/r_h$, where the bump modes have overtaken the ECO ones and designate now the new fundamental mode.} \label{fig:ECO_QNMs+bump_complex_plane_overtaking}
\end{figure*}

Figures \ref{fig:ECO+bump_fund_QNMs} and \ref{fig:ECO+bump_over_QNMs} show collectively the real and imaginary part of the fundamental mode and first overtone for three horizonless ECO configurations with compactness $\mathcal{E}/r_h=10^{-3},\,10^{-2}$ and $10^{-1}$, where the bump location $a_0/r_h$ serves as a free parameter. Firstly, we observe that the fundamental mode of ECOs with compactness $\mathcal{E}/r_h=10^{-3},\,10^{-2}$ remains practically unchanged, regardless of how far (even beyond an astrophysical sense) we place the bump. Thus, compact ($\mathcal{E}/r_h=10^{-2}$) and ultra-compact ($\mathcal{E}/r_h=10^{-3}$) exotic horizonless objects have spectrally-stable fundamental QNMs, for a sufficiently large range of $a_0/r_h$. On the other hand, when the compactness decreases, i.e., the ECO surface is further away from the Schwarzschild radius ($\mathcal{E}/r_h=10^{-1}$), then a migration of the fundamental mode occurs in the complex plane, since beyond $a_0\sim 25 r_h$ both the real and imaginary parts of the fundamental mode change by $\mathcal{O}(10^{-1})$ and $\mathcal{O}(10^{-2})$, respectively. This is a clear occurrence of spectral instability since the lowest-lying mode migrates disproportionately in the complex plane if one considers that the Gaussian bump has an amplitude or order $\mathcal{O}(10^{-6})$\footnote{See though the discussion in Ref. \cite{Boyanov:2024fgc}.}. The existence of these new ``perturbed'' modes, that may destabilize the spectrum, is due to the creation of a secondary trapping cavity in the exterior of the light ring, between it and the environmental bump. We dub these QNMs as \emph{exterior trapped modes}.

\begin{figure*}
    \centering\hspace{-1.25cm}
    \includegraphics[scale=0.6]{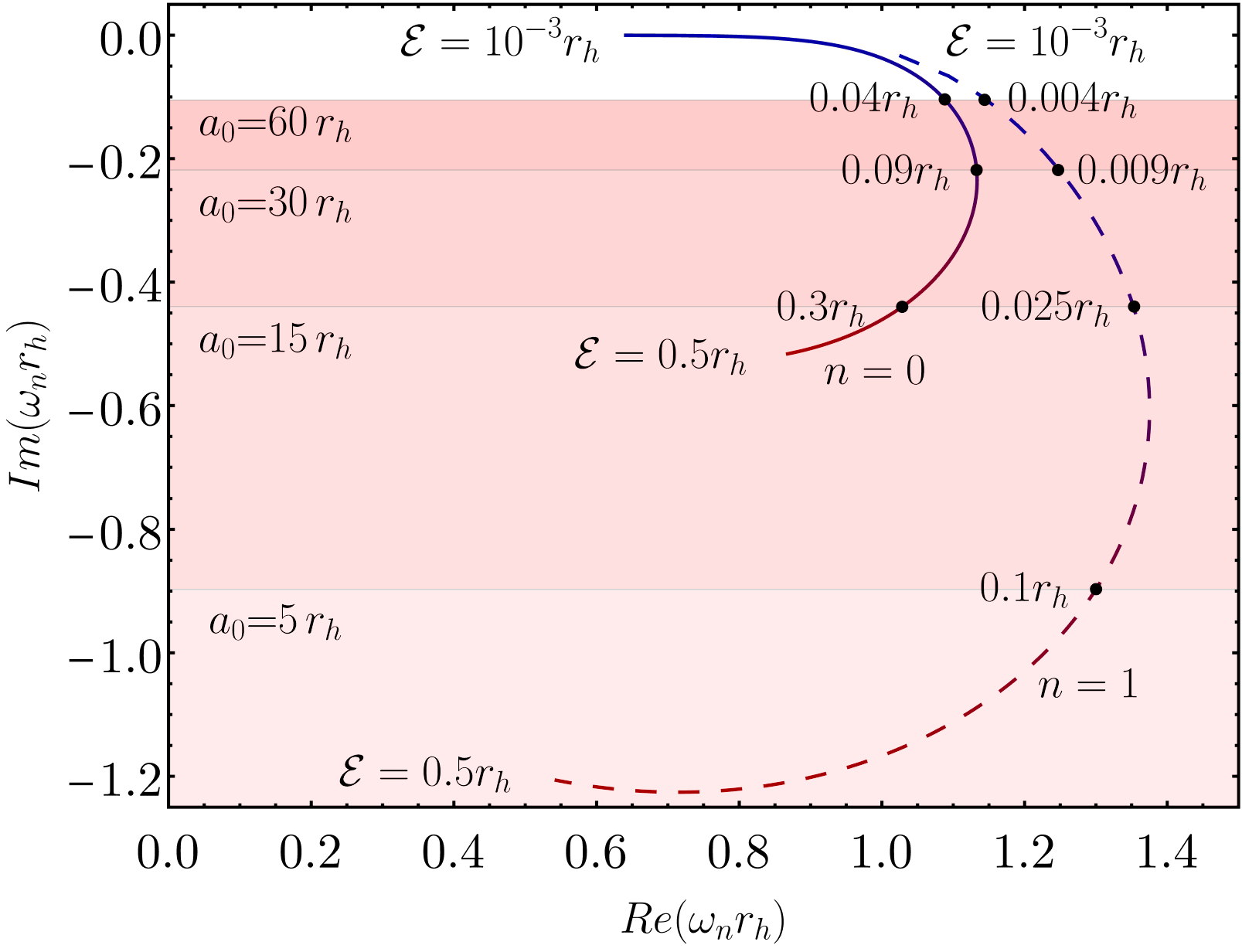}
    \caption{Scalar massless $\ell=2$ QNMs of an ECO, with a perfectly reflective surface at $r=r_s$. By varying the surface position from $\mathcal{E}/r_h=10^{-3}$ to $0.5$ (shown as a gradient from blue to red colors) we plot the evolution of the first two modes, i.e. the fundamental $n=0$ and first overtone $n=1$, in the complex plane. On top of the interior ECO QNMs, we also show with horizontal lines, respective regions of exterior trapped modes between the light ring and the bump location $a_0/r_h$. Different bump positions show different intersection points (shown with black circles) with the interior trapped modes of the ECO as compactness $\mathcal{E}/r_h$ is varied. For example, when the compactness is set to $\mathcal{E}/r_h=0.09$, then if $a_0/r_h\gtrsim30$ the fundamental ECO mode is destabilized. It only takes $\mathcal{E}/r_h=0.009$ for the overtone of the ECO to be destabilized. Thus, it is easier to destabilize a less compact ECO with a small bump than a highly-compact one.} \label{fig:ECO_QNMs+bump_complex_plane_phase_space}
\end{figure*}

We expect that since the fundamental mode of an ECO can be destabilized then the overtones will show equivalent behavior, in accord with pseudospectra analyses \cite{Jaramillo2020}. In Fig. \ref{fig:ECO+bump_over_QNMs} we observe that not only the first overtone of an ECO with $\mathcal{E}/r_h=10^{-1}$ is spectrally unstable, but also the overtone of those with $\mathcal{E}/r_h=10^{-2}$ is destabilized as well, showing an expected yet reassuring behavior. At the same time, we realize that even compact ECOs suffer from spectral instability; the fundamental mode remain stable but the first overtone (and by definition even higher harmonics) shows significant migration. Finally, even though ultra-compact ECOs (with, e.g., $\mathcal{E}/r_h=10^{-3}$) do not show any tendency to spectral instability for the first overtone, there should be some overtone beyond which the spectral instability transpires, as informed by the ECO pseudospectra \cite{Boyanov:2022ark}.

The point of this section though, is not only to find a spectral instability, but also understand if and under what criteria the fundamental mode is destabilized. For that reason, in Fig. \ref{fig:ECO+bump_overtaking} we focus on the case $\mathcal{E}/r_h=10^{-2}$ (left panel) that has a spectrally-unstable $n=1$ overtone, and the case $\mathcal{E}/r_h=10^{-1}$ (right panel) that has a spectrally unstable fundamental mode. On the right panel of Fig. \ref{fig:ECO+bump_overtaking} we show that destabilization of the first overtone and the fundamental mode occurs. In fact, the first overtone is so actively destabilized that after $a_0\sim 28r_h$ it overtakes the ECO fundamental mode and becomes the new \emph{``perturbed'' fundamental mode}, since its imaginary part is the smallest in absolute value. This phenomenon has been dubbed as an \emph{overtaking instability} in Ref. \cite{Cheung:2021bol}.

\subsection{The nature of the overtaking instability}

The case of spectral instabilities, where the overtone takes over the previously considered fundamental mode can in principle occur successively as $a_0/r_h$ increases and various overtakings of modes can take place, as it has been observed in Refs. \cite{Cheung:2021bol,MalatoCorrea:2025iuc}. It is, therefore, important to consider this case more in depth and not just by considering a mode-by-mode proof-of-principle. 

This rather time-consuming but extremely intuitive analysis is presented, firstly, in Fig. \ref{fig:ECO_QNMs+bump_complex_plane_overtaking}. This figure shows (with a solid black curve) the migration of the fundamental interior mode of the ECO with $\mathcal{E}/r_h=10^{-1}$. Besides its outspiral away from the unperturbed fundamental QNM (with $a_0=0$), and the consecutive lessening of its imaginary part (in absolute value), a plenitude of overtones also appear in various shades of red, as well as their respective migration in the complex plane. These red curves are in fact directly associated with the timescales of exterior trapped modes. Finally, the dashed curves represent the position of all modes, including the interior and exterior trapped ones, in the complex plane for fixed bump  positions $a_0/r_h$. It is demonstrated that as long as the bump location is sufficiently close to the light ring (sufficiently close with respect to the particular ECO) the fundamental QNM is the one that originates from the black solid curve that traces the fundamental interior ECO QNM. The rest of the dots designate subdominant exterior trapped overtones of the system. Nevertheless, as we have found, there is a limiting bump location $a_0/r_h$ beyond which the fundamental interior mode is replaced by a perturbed mode. This is now the new ``perturbed'' fundamental mode. This occurs around $a_0/r_h\sim 28$, where the gray-dashed curves become black-dashed. In fact, placing the bump even further, the previously perturbed fundamental QNM is overtaken by other overtones of exterior trapped modes, thus the overtaking instability is a repetitive process, as also found in \cite{Cheung:2021bol,MalatoCorrea:2025iuc}. 

In Fig. \ref{fig:ECO_QNMs+bump_complex_plane_phase_space} we finalize our spectral instability analysis of ECOs within thin matter environments by demonstrating some illustrative parts of QNM destabilization in the phase space $(\mathcal{E},a_0)$, which are in accordance with the results shown in Fig. \ref{fig:ECO_QNMs+bump_complex_plane_overtaking}. We sketch the interior ECO QNMs in the complex plane, as the compactness $\mathcal{E}/r_h$ is varied, where colors transitioning from blue to red designate the increment of $\mathcal{E}/r_h$. We achieve that by placing on top of the curves of the interior fundamental mode and first overtone, the exterior modes found when the bump is placed in a location $a_0/r_h$. Different regions of $a_0/r_h$ are designated with shades of pink. Particular bump locations, are shown with horizontal lines that correspond to the respective imaginary parts of the exterior trapped modes. The intersection of the ECO interior QNMs with the imaginary parts of the exterior ones is shown with black points. 

As Fig. \ref{fig:ECO_QNMs+bump_complex_plane_overtaking} designates that for a fixed compactness, the ECO QNMs are spectrally-destabilized, through an overtaking instability, beyond some bump location, so Fig. \ref{fig:ECO_QNMs+bump_complex_plane_phase_space} shows the phase space ($\mathcal{E}$, $a_0$) where spectral migration and overtaking instability from interior to exterior modes occurs. For example, when $\mathcal{E}/r_h=0.3$ the fundamental mode is overtaken for a bump location at $a_0\gtrsim15r_h$. On the other hand, to destabilize the first overtone of the ECO, with a bump at $a_0\gtrsim15r_h$, a compactness $\mathcal{E}/r_h=0.025$ is only needed. Similarly, when $a_0\gtrsim 60r_h$ then a smaller compactness (larger $r_s$), i.e., $\mathcal{E}/r_h=0.04$ is needed for the fundamental mode to be overtaken and $\mathcal{E}/r_h=0.004$ is needed for the same result on the first overtone. Hence, the more compact the ECO is (smaller surface radius $r=r_s$), the further we need to place the environmental bump from the light ring in order to destabilize the spectrum. Figures \ref{fig:ECO_QNMs+bump_complex_plane_overtaking} and \ref{fig:ECO_QNMs+bump_complex_plane_phase_space} uncover the fact that the QNMs of ultra-compact ECOs present a ``strong'' spectral stability, until they are eventually destabilized with environmental effects, since the bump needs to be located sufficiently far from the light ring of the object. In a similar sense, QNMs from less-compact ECOs (larger surface radius $r=r_s$) can be destabilized ``easier'' by placing a bump in a location closer to the light ring, with respect to ultra-compact configurations. Loosely-compact ECOs with reflective surfaces and a light ring are more prone to spectral instability due to environmental bumps than ultra-compact ones.

The abrupt metamorphosis from a Schwarzschild BH to a horizonless object that we imposed through the reflective boundary condition at the surface, combined with the successive overtaking spectral instability due to environmental effects in non-compact ECOs demonstrate that these objects exhibit both a strong destabilization of their interior, as well as the exterior modes that repetitively takes place, and acquire smaller (in absolute value) imaginary parts with respect to the $a_0=0$ unperturbed QNM. This leads to a series of new ``perturbed'' fundamental QNMs as $a_0/r_h$ is increased. Thus, ECOs present spectral instabilities of both their overtones and fundamental mode, depending on the compactness $\mathcal{E}/r_h$. It is important to note that ultra-compact configurations need radial bump locations that may possibly be too large and, therefore, astrophysically irrelevant. Thus, one could say that the fundamental mode of ultra-compact ECOs cannot be destabilized, in astrophysical scenarios. On the other hand, loosely-compact, purely-reflective, ECOs, even though slightly more astrophysically relevant, present strong destabilization of the
fundamental QNM. Such ECOs are still unrealistic due to their purely reflective surface.

From a thought experiment, where the purely reflective surface transformed slowly to a surface that is partially reflective (and thus also partially absorbing), we may extrapolate that the bump can be placed even further in order to give rise to exterior longer-lived modes that would eventually destabilize the fundamental mode. This is in agreement with the fact that the Schwarzschild BH QNMs can be destabilized completely without the need of near-horizon alterations (see Ref. \cite{Cheung:2021bol}). Thus, the bump is the main source of the spectral instability of QNMs and not the percentage of reflectivity of the ECO surface. Of course, more exploration is needed, where the reflectivity is also altered.

We emphasize that, from a physical standpoint, rotation could in principle destabilize the ECO, potentially invalidating the physical relevance of our modes. However, Ref.~\cite{Maggio:2017ivp} demonstrated that even a small amount of partial absorption is sufficient to stabilize the system. Such absorption can be interpreted as arising from physical mechanisms, for instance effective viscosity (see~\cite{Redondo-Yuste:2025ktt}).

\subsection{Transition from spectral to modal instability?}

An important question arises from the above discussion. Since ultra-compact ECOs possess very long-lived QNMs that are arbitrarily close to the real axis, there might be a way to destabilize them. If they manage to cross the real axis and enter in the upper half of the complex plane, where modally-unstable QNMs reside, then the ECO does not present a stable configuration. This has been suggested by the pseudospectral analysis in \cite{Boyanov:2022ark}, though further non-modal tests in ECOs and Schwarzschild BHs in anti-de Sitter spacetime \cite{Boyanov:2023qqf}, have shown that the protrusion of pseudospectral contour lines on the unstable half of the complex plane, which could be naively interpreted as a transition of a spectral instability to an actual modal instability, is not the case. Not only in Ref. \cite{Boyanov:2022ark}, where the non-modal analysis does not allow for transient phenomena that can destabilize the ECO, but also here, our tentative search for mode destabilization have been proven inconsequential. The fundamental long-lived interior or exterior modes in ultra-compact ECOs are essentially pinned at their spectral points and are not affected by the location or the amplitude (in a logical range of choices) of the bump. We have also studied the $\ell=10$ modes, which show at least three very long-lived QNMs residing arbitrarily close to the real axis \cite{Boyanov:2022ark}, and have obtained identical spectral stability results. Therefore here, we corroborate the non-modal analysis of Ref. \cite{Boyanov:2022ark} with a well-defined theoretical experiment including a horizonless ECO light ring (the ``exotic'' elephant) and a small bump (the flea), which has proven, in other works, to be the key feature of destabilizing the fundamental mode \cite{Jaramillo2020,Destounis:2021lum,Boyanov:2022ark,Courty:2023rxk,Boyanov:2023qqf,Destounis:2023nmb}. 

An important aspect of the mathematical notion of the pseudospectrum \cite{Trefethen:2005} is the fact that it does not only assume theoretical fluctuations to the effective potential of a wave equation, as performed here by hand, but rather perturbs the whole operator that describes wave propagation and evolution. In a sense, the pseudospectra of ECOs shown in Ref. \cite{Boyanov:2022ark} have assumed that not only the effective potential part of the operator has been theoretically perturbed, but also the rest of the parts of the operator are also perturbed. This includes the boundary conditions. Therefore, it might be that the calculation of the pseudospectrum in curved spacetimes can give insights regarding the spectral instability of QNMs but may not have the ultimate ability to predict modal instabilities in marginally-stable, or prone-to-instability spacetime configurations. This issue has only been presented when one deals with curved spacetimes, as far as we know. In fact, there is an abundance of systems that are expected to be stable under a modal (eigenvalue) analysis, though unstable in experiments, with the most famous one being the seminal work in hydrodynamics \cite{Trefethen:1993}, where eigenvalues of a fluid flow predict a stable laminar flow, while the experiment shows transition to turbulent unstable fluid flow. In this case, the pseudospectrum, the transient growth of the fluid operator, and the existence of pseudoresonances perfectly conspire to corroborate the experimental hydrodynamics results. Therefore, we are still a long way from ultimately interpreting the aspects of non-modal analysis in BH spacetimes.

Finally, we note that in this work, we have taken into account idealized, spherically-symmetric ECOs with perfectly reflective surfaces and an astrophysical matter bump. Nevertheless, nature abhors exactly spherically-symmetric configurations. BHs and compact objects in our Universe are rotating, thus are stationary and axially symmetric. These BHs solutions, like the Kerr spacetime, or rotating ECOs \cite{Maggio:2017ivp,Maggio:2018ivz,Maggio:2019zyv,Maggio:2020jml,CardosoPani2019,Zhong:2022jke,Saketh:2024ojw} have more intricate spectra and only a preliminary analysis of the pseudospectrum of Kerr BHs has been achieved, for scalar field perturbations \cite{Cai:2025irl}. Therefore, possible new phenomena that may induce spectral-to-modal instabilities cannot be eliminated till rotation is included to the study performed here. The same holds for the form of the perturbation induced through environmental effects. 

Here, we have used smooth Gaussian bumps, that possess regularity proportional to their sample paths in terms of their derivatives. Perhaps, another Gaussian bump with lower regularity and smoothness might trigger spectral instabilities that lead to the destabilization of the ECO spacetime itself. Nevertheless, for all tests we have performed yet, i.e., increasing the amplitude of the bump $\epsilon$ (up to $\mathcal{O}(10^{-3})$), the distance of the bump from the light ring $a_0/r_h$ (up to almost unrealistic values), and the index $\ell$, that contributes to more modes migrating to the real axis.

\section{Conclusions \& Outlook}

The analysis presented in this work bridges two threads of recent interest in gravitational physics: the spectral instabilities uncovered in the QNM spectrum of BHs \cite{Aguirregabiria:1996zy,Nollert:1996rf,Jaramillo:2022kuv} and horizonless compact objects \cite{Boyanov:2022ark}, and the influence of astrophysical environments on compact-object dynamics \cite{Cole:2022yzw,Cardoso:2021wlq,Cardoso:2022whc,Destounis:2022obl,Speeney:2024mas,Pezzella:2024tkf,Datta:2023zmd,Datta:2025ruh,Gliorio:2025cbh,Spieksma:2024voy,Cheng:2024mgl,Mitra:2025tag,Dyson:2025dlj,Fernandes:2025osu,Destounis:2025tjn}. By modeling the environment as a localized bump \cite{Cheung:2021bol,Courty:2023rxk} outside the light ring of an ECO with perfect reflectivity, we provided a controlled setting to probe the interplay between intrinsic spectral instability and external perturbations.

Our results reveal that the impact of environmental effects is strongly dependent on the ECO compactness. For low-compactness ECOs, the fundamental QNM can be significantly destabilized, underscoring the susceptibility of these configurations to even modest environmental features. By contrast, ultra-compact ECOs display a remarkable robustness: their fundamental QNMs remain spectrally stable despite the presence of the bump. However, this robustness is not universal across the spectrum. The overtones are highly sensitive to environmental perturbations, developing spectral instabilities that alter the QNM hierarchy in a qualitative manner. In particular, we identified an overtaking instability, similar to that in \cite{Cheung:2021bol,MalatoCorrea:2025iuc}, whereby overtones transform into new “perturbed” fundamental QNMs as the environmental bump is displaced away from the light ring. This phenomenon illustrates how spectral instabilities, though not leading to modal divergences \cite{Keir:2014oka}, can restructure the excitation spectrum in ways that would be invisible to a purely modal analysis \cite{Trefethen:1993,Driscoll:1996,Jaramillo:2022kuv,Destounis:2023ruj}.

Furthermore, environmental effects, while able to amplify existing spectral instabilities, are not capable of inducing genuine modal instabilities. This result establishes a clear dichotomy: ECOs can be spectrally fragile yet modally robust, in contrast to typical pseudospectral findings, which are though not in accord with other non-modal analysis tools \cite{Boyanov:2022ark}. Such a distinction is crucial for GW astronomy. Spectral instabilities can leave imprints on the ringdown signal \cite{Jaramillo:2021tmt}, potentially complicating attempts to use BH spectroscopy as a precision test of GR, though their imprints usually manifest themselves in uncharted, late-time territories, where current detectors do not possess the required sensitivity to probe them \cite{Berti:2022xfj,Spieksma:2024voy}.

Our findings demonstrate that the spectra of horizonless compact objects encode a rich interplay between internal structure and external environments. They also emphasize the need to go beyond a strictly modal perspective when assessing the stability and observability of ECOs. Future work should extend this analysis to include rotation, tunable surface reflectivity \cite{Maggio:2021ans}, non-spherical perturbations, and more realistic environmental models, such as accretion flows or dark-matter halo distributions \cite{Cardoso:2021wlq,Courty:2023rxk,Pezzella:2024tkf}, in order to reveal additional dynamical phenomena driven by mode couplings, superradiant amplification, and complex trapping structures.

Ultimately, this work points to a fertile interface between mathematical physics and GW phenomenology. On the theoretical side, we highlight the intricacies in uncovering spectral instabilities that lie beyond the reach of eigenvalue analyses, that is the overtaking instability in which overtones outdistance fundamental modes in the complex plane. On the phenomenological side, spectral instability implies that the ringdown signal of realistic compact objects alters mode hierarchies and excitation patterns that may affect the BH spectroscopy program accordingly. Future detectors such as LISA \cite{LISA:2017pwj,LISA:2022kgy,LISA:2022yao,Karnesis:2022vdp} and third-generation ground-based interferometers \cite{Punturo:2010zz,ET:2019dnz,Reitze:2019iox,Hall:2022dik}, with their enhanced sensitivity to subtle spectral features, may provide the observational reach needed to test these predictions. By integrating refined theoretical tools with GW data analysis, it might become possible not only to constrain or identify ECOs but also to sharpen the role of the ringdown stage as a probe of the event horizon, as well as new physics in the strong-gravity regime.

\begin{acknowledgments}
K.D. acknowledges financial support provided by FCT – Fundação para a Ciência e a Tecnologia, I.P., under the Scientific Employment Stimulus – Individual Call – Grant No. 2023.07417.CEECIND/CP2830/CT0008. K.D. acknowledges financial support provided by FCT–Fundação para a Ciência e a Tecnologia, I.P., under the Scientific Employment Stimulus – Individual Call – Grant No. 2023.07417.CEECIND/CP2830/CT0008. K.D. would also like to thank the Fundação para a Ciência e Tecnologia (FCT), Portugal, for the financial support to the Center for Astrophysics and Gravitation (CENTRA/IST/ULisboa) through grant No. UID/PRR/00099/2025 and grant No. UID/00099/2025.
M.M.C. and C.F.B.M. acknowledge Fundação Amazônia de Amparo a Estudos e Pesquisas (FAPESPA), Conselho Nacional de Desenvolvimento Científico e Tecnológico (CNPq) and Coordenação de Aperfeiçoamento de Pessoal de Nível Superior (CAPES) – Finance Code 001, from Brazil, for partial financial support.
R.P.M. acknowledges support from the Villum Investigator program supported by the VILLUM Foundation (grant no. VIL37766) and the DNRF Chair program (grant no. DNRF162) by the Danish National Research Foundation. The Center of Gravity is a Center of Excellence funded by the Danish National Research Foundation under grant No. 184.
This project has received funding from the European Union’s Horizon MSCA-2022 research and innovation programme “Einstein Waves” under grant agreement No. 101131233.
\end{acknowledgments}

\bibliography{biblio}

\end{document}